\documentclass[showpacs,preprintnumbers,amsmath,amssymb]{revtex4}

\usepackage{graphicx}% Include figure files
\usepackage{subfigure}
\usepackage{dcolumn}% Align table columns on decimal point
\usepackage{bm}% bold math

\begin{document}

\preprint{APS/123-QED}

\title{Role of an intermediate state in homogeneous nucleation}% Force line breaks with \\

\author{Takaaki Monnai}%
\email{monna@suou.waseda.jp}%
\author{Ayumu Sugita}
 \email{sugita@a-phys.eng.osaka-cu.ac.jp}
\author{Katsuhiro Nakamura}
\email{nakamura@a-phy.eng.osaka-cu.ac.jp}
\affiliation{$*$Department of Applied Physics ,Waseda University, 3-4-1 Okubo,
Shinjuku-ku, Tokyo 169-8555, Japan\\
$\dagger $$\ddagger $Department of Applied Physics, Osaka City University, 
3-3-138 Sugimoto, Sumiyoshi-ku, Osaka 558-8585, Japan}%Lines break automatically or can be forced with \\

\date{\today}% It is always \today, today,

\begin{abstract}
We explore the role of an intermediate state (phase) 
in homogeneous nucleation 
phenomenon by examining the  decay process through 
a doubly-humped potential 
barrier.  
As a generic model we use the fourth- and sixth-order Landau potentials 
and analyze the Fokker-Planck equation for the one-dimensional 
thermal diffusion in the system characterized by a triple-well potential.
In the low temperature case we apply the WKB method to the decay process
and obtain the decay rate which 
is accurate for a wide range of depth and curvature of the middle well.
In the case of a deep middle well, 
it reduces to a doubly-humped-barrier counterpart 
of the Kramers escape rate:
the barrier height and the 
curvature of an initial well in the Kramers rate are replaced by 
the arithmetic mean of higher(or outer) and 
lower(or inner) partial barriers
and the geometric
mean of curvatures of the initial and intermediate wells, respectively.
It seems to be a universal formula.   
In the case  of a shallow-enough middle well, Kramers 
escape rate is alternatively evaluated within the standard framework of 
the mean-first-passage time problem, which certainly supports the WKB result. 
The criteria whether or not the existence 
of an intermediate state can enhance 
the decay rate are revealed.
\end{abstract}

\pacs{05.40.-a, 05.45.Mt, 82.20.Db}% PACS, the Physics and Astronomy
\maketitle

\section{\label{introduction}INTRODUCTION}
The decay process for a metastable system 
received considerable attention 
for long time.
By using the WKB analysis of the Fokker-Planck (FP) equation, 
van Kampen\cite{VanKampen} investigated 
the role of fluctuations in thermal diffusion process 
under a symmetric bistable system and found that 
the system's ergodicity enables a 
Brownian particle to escape over the potential barrier. 
In asymptotic time regime, he also obtained 
the astonishingly accurate formula 
for the so-called Kramers escape rate, i.e., the rate at which 
a Brownian particle escapes from a potential well 
over the single-humped barrier.
Later, Caroli et al.\cite{Caroli} explored 
the asymmetric bistable system using a path 
integral (instanton) formalism.  
The separation of the time scale required 
for the Kramers rate treatment was investigated by Kapral\cite{Kapral} 
by using the projection operator approach. For a Langevin equation with 
spatially- and time-correlated colored noise in the 
case of weak damping, the staggered-ladder spectra of 
the corresponding FP equation are found\cite{Arvedson}. 

Recently, Nicolis and co-workers\cite{Nicolis,NicolisBasios} 
studied the decay in the system with 
a metastable intermediate state in the
context of the protein crystallization, 
and within the rate equation treatment 
they gave an important suggestion that
the existence of the intermediate state 
can enhance the nucleation rate under some parameter range.
We note that one-dimensional treatment of chemical reaction 
along the reaction coordinate in the free energy landscape is very 
useful to describe the above problem. 

In this paper, we investigate the role of 
an intermediate state in nucleation phenomenon based 
on the WKB analysis of FP equation 
for the one-dimensional order parameter, and present 
a doubly-humped-barrier counterpart of 
the Kramers rate\cite{Kramers,Frenkel}.  
The decay rate is given as the 1st excited 
eigenvalue of the FP operator.
As is well known, FP equation is transformed into the associated 
Schr\"odinger equation and in the case that
a well-defined thermal equilibrium does exist, 
eigenvalues of FP equation are exactly the same as those of 
Schr\"odinger equation\cite{VanKampen,Sasada,Risken}. Further, 
in the low-temperature case we can apply 
the WKB analysis to the system with
a triple-well potential and extend Kramers escape 
rate to the situation with 
a doubly-humped barrier.

This paper is organized as follows.
In Section 2 we reduce the FP equation into 
the Schr\"odinger equation and
obtain the WKB quantization condition for a system 
with triple-well potential. 
Section 3 is concerned with the evaluation 
of the tunneling integral, 
leading to
the extended Kramers formula for 
the doubly-humped potential barrier. 
In Section 4 the condition for the enhancement 
of nucleation rate caused by the intermediate state is revealed. 
In Section 5 we numerically 
check the validity of our new formula for the nucleation rate.
Appendix is devoted to the alternative 
evaluation of the nucleation rate in the case of a shallow middle well,
within the standard framework of the mean-first-passage
time problem. 

\section{\label{WKB}Fokker-Planck equation and WKB analysis}
Relaxation process responsible for a homogeneous nucleation is described by FP equation as
\begin{equation}
\frac{\partial}{\partial t}P(x,t)=
\frac{1}{\eta}\frac{\partial}{\partial x} 
\left(\frac{\partial W(x)}{\partial x}+\theta \frac
{\partial}{\partial x}\right)P(x,t).
\end{equation}
$\eta$ is the viscosity and $\theta=k_B T$ is the temperature. 
Hereafter we assume $\eta=1$ for simplicity.
In order to specify the problem, we consider the symmetric 
triple-well potential
proposed by Nicolis {\it et al.} 
\cite{Nicolis}, 
\begin{equation}
W(x)=\int dx x (x^2-\lambda) (x^2-\mu)
\end{equation}
with $\lambda>\mu>0$,
which has two potential minima at $x=\pm \sqrt{\lambda}$, 
two maxima at $x=\pm \sqrt{\mu}$ 
and one local minimum at $x=0$ as shown 
in Fig.1(a). This potential mimics the barrier 
accompanied with a cave on the
top of the double-well potential, which 
represents the presence of the intermediate state. 
By the separation anzats 
$P(x,t)=e^{-\frac{W}{2\theta}}\phi(x)e^{-E t}$, the FP equation is 
transformed into the associated Schr\"odinger equation, 
\begin{equation}
H\phi(x)= -\theta \phi^{''}(x) + V(x)\phi(x) = E\phi(x),
\label{assoSch}
\end{equation}
with the effective potential 
\begin{equation}
V(x)=\frac{W'^2}{4\theta}-\frac{W''}{2}
\end{equation}
and $\theta$ corresponding to 
$\theta=\frac{\hbar^2}{2m}$ in quantum mechanics 
with Planck constant 
$\hbar$ and mass $m$. Equation (\ref{assoSch}) has 
the self-evident lowest-eigenvalue $E_0=0$ with the eigenfunction 
$\phi_0(x)=e^{-\frac{W}{2\theta}}$ and the remaining eigenvalues 
satisfy $E_n>E_0$.  The problem of the time evolution of 
the probability density $P(x,t)$
initially distributed in the left well reduces to 
that of finding the lowest 
few eigenvalues of Eq.(\ref{assoSch}). 
Eventually the decay rate $\Gamma$ 
is given by the first-excited eigenvalue.

As shown in Fig.1(b), $V(x)$ has also symmetric 
three wells and two barriers. 
We note: even if the central well is shallow in the original potential 
$W(x)$ (see Fig.2(a)), the effective potential $V(x)$ is negative at the  
corresponding local minimum ($V(0)= -\frac{W''(0)}{2}<0$),  
and for the original triple-well potential the effective potential 
$V(x)$ has three potential minima, all with negative values (Fig.2(b)). 
Thus {\it the lowest three eigenvalues of Eq.(\ref{assoSch}) 
including $E_0=0$ experience all 
the three potential minima and two maxima} (see Figs.1(b) and 2(b)). 
We are particularly interested in the situation where 
the lowest eigenvalues in each of three wells are almost degenerate 
without tunneling through barriers. Then, by applying the WKB method 
we shall 
remove the degeneracy by incorporating the tunneling effects.  

Let us denote the regions bordered by the classical 
turning points $x_0,\cdots,x_5$ as $a,b,c,d$, and $e$ (see Fig.1(b)). 
\begin{figure} \center{
\includegraphics[scale=0.6]{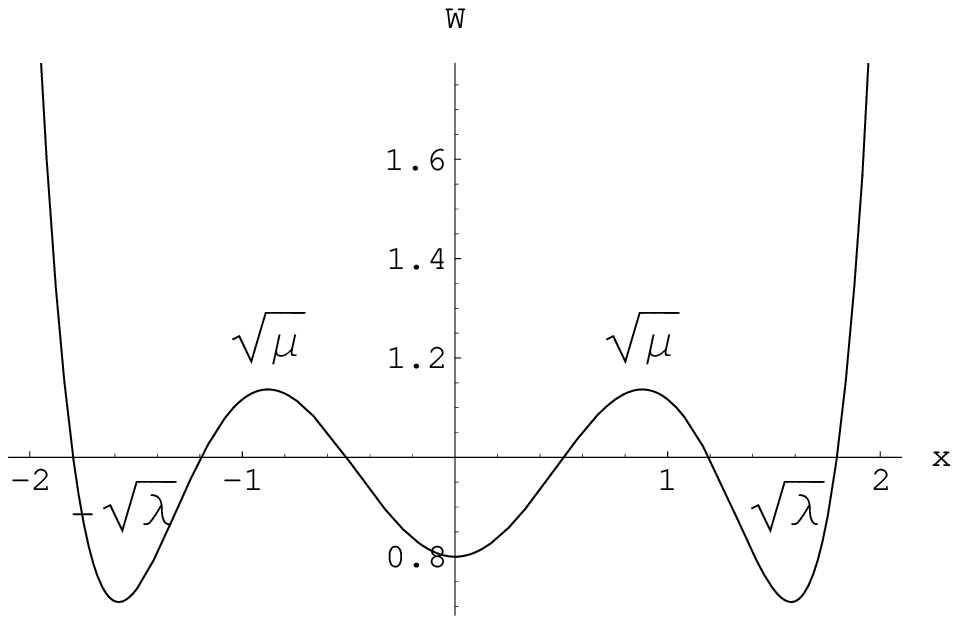}
\includegraphics[scale=0.6]{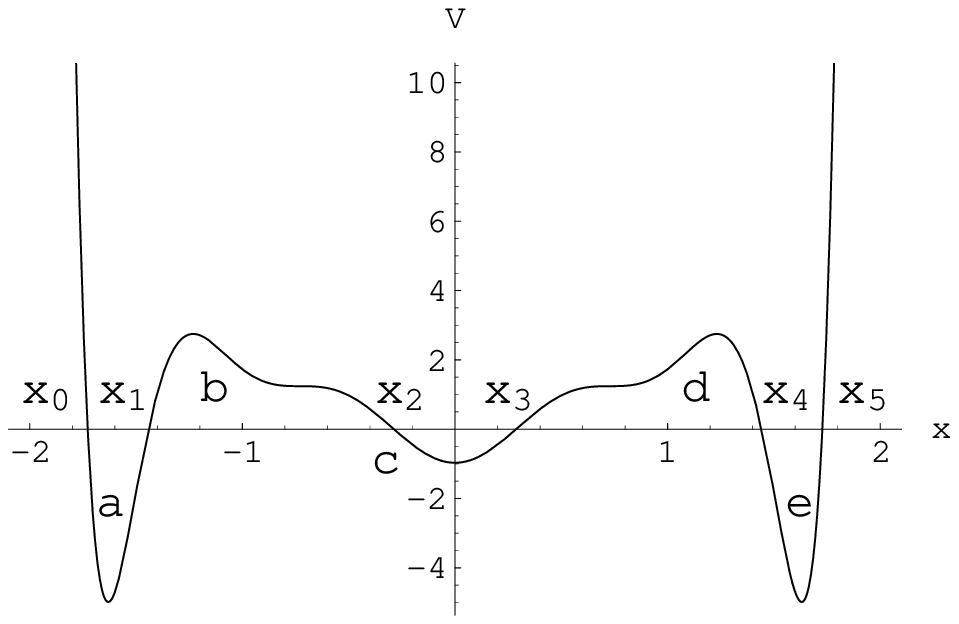} }
\caption{(a) Landau potential $W(x)$ in the presence 
of intermediate state in 
the case of deep central cave with $\lambda=2.5,\mu=0.775$; 
(b)Effective potential $V(x)$ corresponding to Fig.1 (a).}
\end{figure}
\begin{figure}
\center{
\includegraphics[scale=0.6]{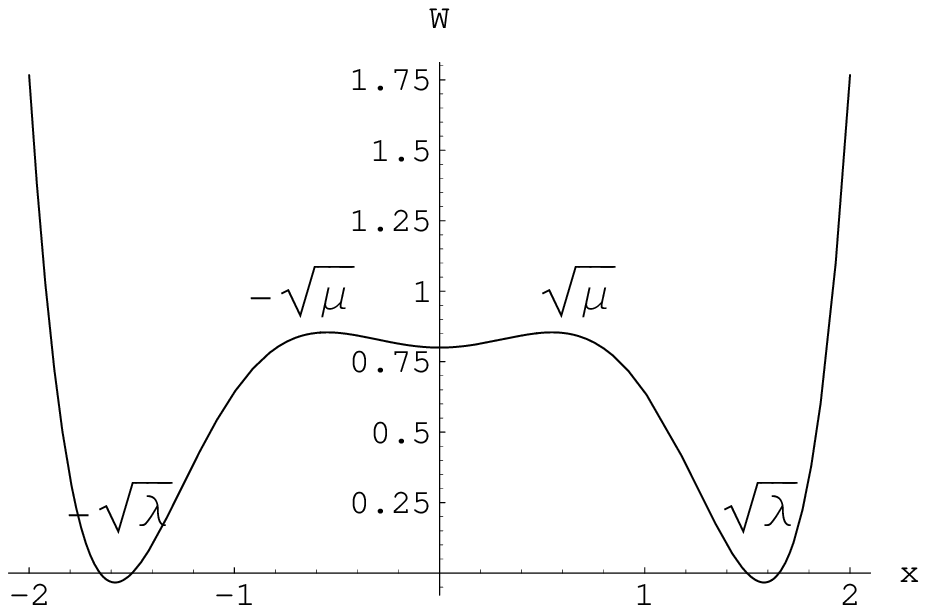}
\includegraphics[scale=0.6]{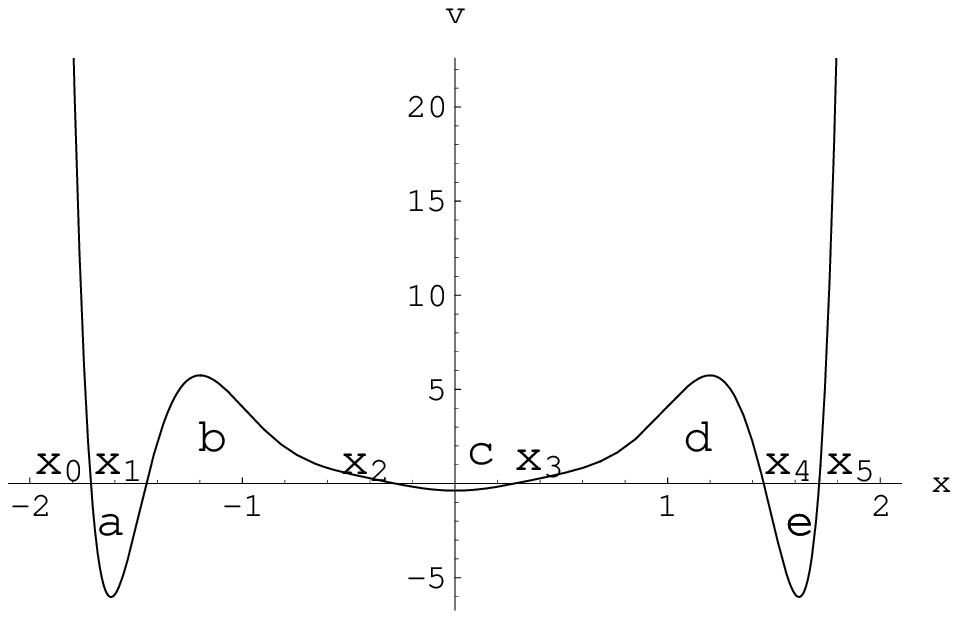} }
\caption{(a)Landau potential $W(x)$ in the presence of intermediate state 
in the case of shallow central cave with $\lambda=2.5,\mu=0.3$.
(b)Effective potential $V(x)$ corresponding to Fig.2 (a). 
Note: although $W(x)$ has a shallow central cave, 
the bottom of the middle well of $V$ is negative unlike the case 
of double-well potential with a single-humped barrier.}
\label{Figg}
\end{figure}
The wave functions in each regions $\phi_a,\cdots,\phi_e$ 
are connected by the connection formula, 

\begin{eqnarray}
&&\frac{1}{\sqrt{p}}e^{\pm i(S+\frac{\pi}{4})}\quad (E>V)
\longleftrightarrow 
\frac{1}{\sqrt{p}}(e^S\pm\frac{i}{2}e^{-S})\quad (E<V)\nonumber \\
&&S=|\frac{1}{\hbar}\int_{y_0}^x p dx|, \quad p=\sqrt{2m |E-V(x)|} 
\end{eqnarray}
where $y_0$ denotes a classical turning point.
The wave function in the left-most region $a$, is 
written as $\phi_a=\frac{c}{\sqrt{p}}
\sin(\frac{1}{\hbar}\int_{x_0}^{x}p dx
+\frac{\pi}{4})$, since it
does not contain a term which exponentially diverges in the region $-\infty < x <x_0$.
By using the connection formula, we can successively 
connect the wave functions until  $\phi_e$ in the right-most region as
\begin{eqnarray}
&\phi_e&=\frac{c}{\sqrt{p}}\{(-2i\cos S_a\cos S_c 
e^{M_b+M_d}+\frac{i}{2}\sin S_a \sin S_c e^{-M_b+M_d}
\nonumber \\
&&+\frac{1}{2}\cos S_a\sin S_c e^{M_b-M_d}+ 
\frac{1}{8}\sin S_a\cos S_c e^{-M_b-M_d})
e^{iS_e}e^{-\frac{i}{\hbar}\int_{x}^{x_5}p dx} e^{i\frac{\pi}{4}}
\nonumber \\
&&+(2i\cos S_a\cos S_c e^{M_b+M_d}-\frac{i}{2}
\sin S_a \sin S_c e^{-M_b+M_d}+\frac{1}{2}
\cos S_a\sin S_c e^{M_b-M_d}\nonumber \\
&&+\frac{1}{8}\sin S_a\cos S_c e^{-M_b-M_d})
e^{-iS_e}e^{\frac{i}{\hbar}\int_{x}^{x_5}p dx} e^{-i\frac{\pi}{4}}\}.
\end{eqnarray}
$S_{\alpha}$ and $M_{\beta}$ are the action and tunneling integrals in each well $\alpha=a,c,e$ and barrier $\beta=b,d$, respectively. 
Hereafter, wherever necessary, we use the abbreviated notation $\alpha=a,c,e$ to indicate the wells $a,c$, 
and $e$, respectively.     
Since $\phi_e$ in the right-most well should not contain a term, $\sin(\frac{1}{\hbar}
\int_x^{x_5}p dx-\frac{\pi}{4})$, connecting to the term which exponentially diverges in the
region  $x_5< x <\infty$, 
we obtain the following relation as \cite{Sasada,OhtaNakamura}
\begin{equation}
4 \cot S_a \cot S_c\cot S_e-\cot S_e e^{-2M_b}-\cot S_a 
e^{-2 M_d}-\frac{1}{4}\cot S_c e^{-2(M_b+M_d)}=0.\label{quantization}
\end{equation}
Due to the small transparency of each barrier, 
actions are approximately quantized as $S_{\alpha}\approx \pi (n_{\alpha}+\frac{1}{2})$
and one sees 
\begin{equation}
\cot S_{\alpha}\approx \pi \left(n_{\alpha}+\frac{1}{2}-S_{\alpha}\right).\label{cot}
\end{equation}
Since  our  interest lies in the low-lying energy levels, we consider the case $n_a=n_c=n_e=0$.
In this case, for low-lying eigenvalue $E$, action is evaluated as 
$S_{\alpha}(E)\approx (E-E_{\alpha})\frac{dS}{dE}{\large \mid }_{E=E_\alpha}+S(E_{\alpha})=\pi\frac{E-V_{\alpha}}{\hbar \omega_{\alpha}}$
and substituting (\ref{cot}) into (\ref{quantization}), one 
obtains the equation in the form of cubic polynomials of eigenvalue $E$, 

\begin{eqnarray}
&&4\pi^3\left(\frac{E-V_{a}}{\hbar \omega_a}-\frac{1}{2}\right)
\left(\frac{E-V_{c}}{\hbar \omega_c}-\frac{1}{2}\right)
\left(\frac{E-V_{e}}{\hbar \omega_e}-\frac{1}{2}\right)\nonumber \\
&&\quad -\pi\left(\frac{E-V_a}{\hbar \omega_a}-\frac{1}{2}\right)
e^{-2 M_d}-\pi
\left(\frac{E-V_{e}}{\hbar \omega_e}-\frac{1}{2}\right)
e^{-2 M_b}\nonumber \\
&&\qquad-\frac{1}{4}\pi
\left(\frac{E-V_{c}}{\hbar \omega_c}-\frac{1}{2}\right)
e^{-2 (M_b+M_d)}=0.
\label{energy}
\end{eqnarray}
Here
$E_{\alpha}$, $V_{\alpha}\equiv (E_{\alpha}-\frac{\hbar\omega_{\alpha}}{2})$ and 
$\omega_{\alpha}\equiv\frac{\pi}{\hbar}(\frac{dS}{dE}{\large{\mid }}_{E=E_{\alpha}})^{-1}$
defined for each well $\alpha$ imply the "vacuum energy", (approximate) potential 
minimum 
and the frequency, respectively.
\section{\label{tunneling}tunneling integrals and decay rate}
The precise solutions of Eq.(\ref{energy}) 
constitute the lowest three eigenvalues with 
the lowest one being the ground state energy $E_0=0$.   
To reach the goal,  the evaluation of tunneling  integrals for each well is necessary.
In the problem under consideration, the potential is assumed symmetric 
and thus $V_a=V_e$, $M_b=M_d$, and $\omega_a=\omega_e$.
The tunneling integrals are perturbatively estimated 
with use of small parameter $\theta$.
The classical turning points gives the 
interval $[x_3,x_4]$ on which the tunnel integration is performed.
As shown in Figs.1 and 2, the  interval 
contains only one local maximum  at $x=y$ of the original potential $W$.
In this case the tunnel integral is calculated as follows:
We divide the integral
\begin{equation}
\theta M_d=\int_{x_3}^{x_4}\sqrt{\frac{W'^2}{4}-\frac{\theta W''}{2}} dx
\end{equation}
into three parts and perform the integration as
\begin{eqnarray}
&&\theta M_d=\int_{x_3}^{y-\Delta}
\frac{W'}{2}(1-\theta \frac{W''}{W'^2}+O(\theta^2))dx\nonumber \\
&&+\int_{y-\Delta}^{y+\Delta}
\sqrt{\frac{-\theta W''}{2}}(1-\frac{W'^2}{4\theta W''}+
O(\left(-\frac{W'^2}{2\theta W''}\right)^2 ))dx-
\int_{y+\Delta}^{x_4}\frac{W'}{2}(1-\theta \frac{W''}{W'^2}+
O(\theta^2))dx\nonumber \\
&&=-\frac{W(x_3)+W(x_4)}{2}+W(y)-\theta \log(-W''(y)\sqrt{\theta})+O(\theta).
\end{eqnarray}
We choose $\Delta =\sqrt{|\frac{2\theta}{W''(y)}|}$ 
so that the above-used Taylor expansions,
$\sqrt{1+x}=1+\frac{x}{2}+O(x^2)$ are possible.

\begin{figure}
\center{
\includegraphics[scale=0.6]{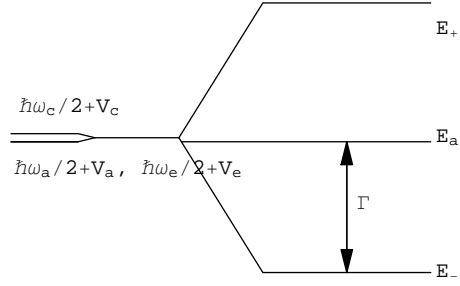}
}
\caption{Energy splitting and shifts in the case (i) of a deep middle well.
The unperturbed three eigenvalues are almost degenerate.}
\label{levelsplit}
\end{figure}

\begin{figure}
\center{
\includegraphics[scale=0.6]{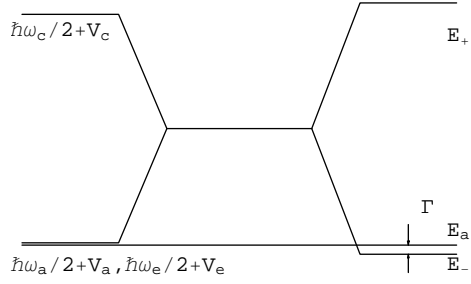}
}
\caption{Energy splitting and shifts 
in the case (ii) of a shallow middle well.
The unperturbed eigenvalue for a central well is separated from the
doubly-degenerate ones in the left-most and right-most wells.}
\label{levelsplit2}
\end{figure}

Equation (\ref{energy}) is solved as follows:
When the barrier heights are large enough, 
the terms including the exponentially-small tunneling factors are negligible 
on the left-hand side of (\ref{energy}), 
giving a set of the lowest eigenvalues at each of separate wells, 
$E_{a,e}=\frac{\hbar\omega_a}{2}+V_a$ 
and $E_c=\frac{\hbar\omega_c}{2}+V_c$.
When the effect of the small transparency 
due to tunneling will be incorporated,
these three levels show a splitting and 
shifts (see Figs. \ref{levelsplit} and
\ref{levelsplit2}). 

In such a case one can as yet suppress the
fourth term on the left-hand side of (\ref{energy}), 
which is a product of two tunneling terms and 
is almost vanishing, and one obtains the exact eigenvalues:
\begin{eqnarray}
&&E_a=\frac{\hbar\omega_a}{2}+V_a,\nonumber \\
&&E_{\pm }= \frac{\frac{\hbar\omega_a}{2}+
V_a+ \frac{\hbar\omega_c}{2}+V_c}{2} \nonumber \\
&& \qquad \pm \frac{1}{2}
\sqrt{((\frac{\hbar\omega_a}{2}+V_a)-
(\frac{\hbar\omega_c}{2}+V_c))^2+
\frac{2\hbar^2\omega_a\omega_c}{\pi^2}e^{-2M_d}},
\label{exact-sol}
\end{eqnarray}
where $E_{-}$ stands for $E_0=0$.
The decay rate is given as the difference 
of two low-lying eigenvalues, 
$\Gamma =E_a-E_-$.

Depending on the depth of the middle well of $V(x)$, 
there are two typical cases,\\
\noindent
case (a): $|(\frac{\hbar\omega_a}{2}+V_a)-
(\frac{\hbar\omega_c}{2}+V_c)| 
\ll \frac{\sqrt{2}\hbar\sqrt{\omega_a\omega_c}}{\pi}e^{-M_d}$;\\
\noindent
case (b): $|(\frac{\hbar\omega_a}{2}+V_a)-
(\frac{\hbar\omega_c}{2}+V_c)| \gg 
\frac{\sqrt{2}\hbar\sqrt{\omega_a\omega_c}}{\pi}e^{-M_d}$.\\
\noindent

The cases (a) and (b) corresponds to deep and shallow middle wells, 
respectively. The energy splitting and shifts in each 
of these cases are illustrated 
in Figs. \ref{levelsplit} and \ref{levelsplit2}. 
Our principal interest lies in the case (a) that 
the unperturbed three levels without 
tunneling terms are almost degenerate. 
In this case the first term in the square 
root in Eq.(\ref{exact-sol}) is 
negligible and
the decay rate is dominated by the tunneling term as

\begin{equation}
\Gamma=E_a-E_-=\sqrt{\frac{4\hbar^2\omega_a\omega_c\theta|W''(y)|}
{\pi^2|
W'(x_4)W'(x_3)|}}e^{1-
\sqrt{2}-Arcsihh(1)}e^{-\frac{W(y)-W(x_3)+
W(y)-W(x_4)}{2\theta}}
\label{tunnel-a}
\end{equation}
with the turning points $x_3=\sqrt{\frac{2\theta}{\lambda\mu}}, 
x_4=\sqrt{\lambda}-\sqrt{\frac{\theta}{\lambda(\lambda-\mu)}}$
and the maximum point $y=\sqrt{\mu}$ in the tunneling integration.

\begin{figure}
\center{
\includegraphics[scale=0.5]{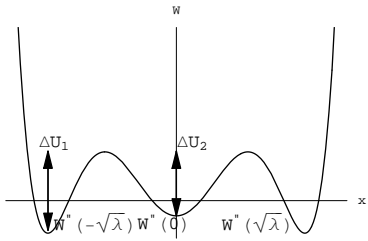}
}
\caption{The potential barriers 
$\Delta U_1$ and $\Delta U_2$  and curvatures 
in the original potential $W(x)$.}
\label{meaning}
\end{figure}

Using the harmonic approximation $\hbar\omega_{a,c,e}\approx\sqrt{2\theta V''(x_{a,c,e}^{min})}$
for the curvature of each well of $V(x)$, (\ref{tunnel-a}) is rewritten in the lowest order 
of $\theta$ as 
\begin{equation}
\Gamma=\frac{\sqrt{2}-1}{\pi}e^{2-\sqrt{2}}
\sqrt{2\sqrt{W''(0)W''(\sqrt{\lambda})}
|W''(\sqrt{\mu})|}e^{-(\Delta U_1 +\Delta U_2)/2\theta}
\label{WKB}
\end{equation}
where 
$\Delta U_1=
W(\sqrt{\mu})-W(\sqrt{\lambda})$ and 
$\Delta U_2=W(\sqrt{\mu})-W(0)$ 
are the height of potential barriers 
at the top measured from the bottom of 
the left-most well and from the bottom of the middle cave, respectively.
$W''(0), W''(\sqrt{\mu})$ and $W''(\sqrt{\lambda})$ 
are the curvatures 
at the cave, at the barrier top and 
at the the bottom of the left-most well, 
respectively. 
 These quantities are illustrated in Fig.\ref{meaning}.
In deriving (\ref{WKB}) from (\ref{tunnel-a}) we used the expressions valid in the lowest order 
of $\theta$, $\sqrt{2\theta V''(x_{a,e})}=W''(\sqrt{\lambda})$, $\sqrt{2\theta V''(x_{c})}=W''(0)$,
$W'(x_3)=\sqrt{2\theta W''(0)}$ and  $W'(x_4)=\sqrt{2\theta W''(\sqrt{\lambda})}$.

The result in Eq.(\ref{WKB}) 
is a doubly-humped-barrier counterpart of the Kramers escape rate 
for a single barrier, since 
{\it  the potential-barrier height and the 
curvatures of a well in Kramers rate are replaced 
by 
the arithmetic mean of higher(or outer) and 
lower(or inner) partial barriers ($\Delta U_1$ and $\Delta U_2$)
and the geometric
mean of curvatures of 
the initial and intermediate wells 
($W''(\sqrt{\lambda})$ and $W''(0)$), 
respectively} (see Fig.\ref{meaning}). 
 The result in Eq.(\ref{WKB})
cannot be obtained within the standard 
framework of the mean-first-passage 
time problem.  

On the other hand, in the case (b), 
one can Taylor-expand the formula 
in Eq.(\ref{exact-sol}) in the small tunneling term, finding
$E_{-}=E_a-\frac{1}{2|(\frac{\hbar\omega_a}{2}+V_a)-
(\frac{\hbar\omega_c}{2}+V_c)|}
\frac{\hbar^2\omega_a\omega_c}{\pi^2}
e^{-2M_d}$. In this case the decay rate is given by
\begin{equation}
\Gamma=E_a-E_{-}=
\frac{(\sqrt{2}-1)^2}{\pi^2}e^{4-2\sqrt{2}}
\frac{2\sqrt{W''(0)W''(\sqrt{\lambda})}|W''(\sqrt{\mu})|}
{|(\frac{\hbar\omega_a}{2}+V_a)-(\frac{\hbar\omega_c}{2}+V_c)|}
e^{-\frac{\Delta U_1 +\Delta U_2}{\theta}}.
\label{Kramers}
\end{equation}
Although in the case of a shallow middle well the difference
of harmonic energy levels can not 
be written in terms of original potential,
the activation energy,
$\Delta U_1 +\Delta U_2$ with $\Delta U_2 \approx 0$ 
is nearly equal to 
the barrier height at the barrier top measured
from the bottom of the left-most well 
and the conventional Kramers escape rate is recovered. 
As is proved in Appendix, 
the result 
in Eq.(\ref{Kramers}) can be confirmed by alternatively solving
the corresponding rate within 
the standard framework of the mean-first-passage 
time problem. 

\section{Enhancement of nucleation rate}\label{enhancement}
We proceed to investigate the possibility of 
enhancement of nucleation process due to 
the existence of an intermediate state
on the basis of the low-temperature formula 
for the doubly-humped barrier 
version of Kramers rate.

\begin{figure}
\center{
\includegraphics[scale=0.5]{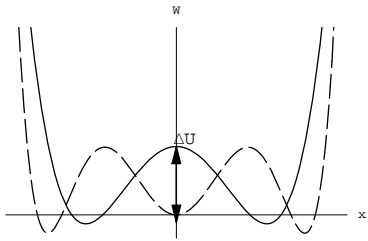}
}
\caption{Schematic symmetric double-well potential $W(x)$
with (broken line) and without (solid line) 
a cave in the central barrier.
$\Delta U$ implies a barrier height
in the case of the double-well potential with no cave
in the barrier.}
\label{Figg6}
\end{figure}

In the case of a symmetric double-well potential in Fig.\ref{Figg6}, 
$W(x)=\frac{x^4}{4}-\frac{1}{2}\nu x^2$, the decay rate 
is calculated \cite{VanKampen} in the same way as 
in the previous Section and is given by
\begin{equation}
\Gamma'=\frac{\sqrt{2}-1}{\pi}\sqrt{W''(\sqrt{\nu})|W''(0)|}
e^{2-\sqrt{2}}e^{-\frac{\Delta U}{\theta}}
\end{equation}
with a potential barrier $\Delta U=W(0)-W(\sqrt{\nu})$ 
(see Fig.\ref{Figg6}).
We note that a faster decay due to 
the existence of an intermediate state is 
guaranteed
under the following conditions:\\
\noindent
(i) the Boltzmann factor for 
a doubly-humped barriers is larger than that 
for a single barrier;\\
\noindent
(ii) Provided that two kind of Boltzmann factors are identical, 
the pre-exponential factor for 
a doubly-humped barriers is larger than that 
for a single barrier.\\
\noindent 

The condition (i) implies that
the mean barrier height $\frac{\Delta U_1+\Delta U_2}{2} $ 
for triple wells is less than the single-barrier 
height for the double-well $\Delta U$.
To be explicit, the condition (i) is given by
\begin{equation}
\nu > \sqrt{G(\lambda,\mu)}, \qquad
G(\lambda,\mu)\equiv \frac{\lambda^3-
3\lambda^2\mu+6\lambda\mu^2-2\mu^2}{6}.
\end{equation} 

On the other hand, in the case (ii) the quasi-equality
$\frac{\Delta U_1+\Delta U_2}{2}\approx\Delta U$ leads to
\begin{equation}
\nu \approx \sqrt{G(\lambda,\mu)}.
\label{quasi-eq}
\end{equation} 
Namely, $\nu$ is almost a function of $\lambda$ and $\mu$.
For the pre-exponential 
factors to satisfy $ \Gamma>\Gamma'$, we have 
\begin{equation}
\lambda \mu^{\frac{3}{2}}(\lambda-\mu)^{\frac{3}{2}}>2^{-3/2}\nu^2. 
\label{pr-exp}
\end{equation}
Equations (\ref{quasi-eq}) and (\ref{pr-exp}) 
satisfy the condition (ii).

In the same way, enhancement of the nucleation 
due to the intermediate state is also expected in the case of 
the sixth-order double-well Landau potential, 
$W(x)=\frac{x^6}{6}-\frac{\kappa}{4}x^4$. 
For instance, the condition (i) is now given by 
\begin{equation}
\kappa > (3G(\lambda,\mu))^{\frac{1}{3}}.
\end{equation}
\section{Numerical evaluation of decay rate}
To verify the results in the previous Sections, we shall
investigate 
the decay rate numerically by
calculating the second-lowest eigenvalue of 
the associated Schr\"odinger operator 
$H=\frac{p^2}{2}+V(x)$ 
in Eq.(\ref{assoSch}) \cite{Dekker,Monnai}.
By using the eigenvectors of the harmonic oscillator together 
with the annihilation-creation operator 
representation of $p$ and $x$, 
${\langle n|H|m\rangle}$ is diagonalized 
straightforwardly \cite{Tissen}. 
For the system with a triple-well potential, the decay rate 
is obtained numerically in the parameter range 
$\theta=0.1$, $2<\lambda<3,\quad 0.3<\mu<1$, which is
compared with the formula in Eq.(\ref{WKB}) in
the case of a deep middle well (Fig.7).  
In Fig.7(a), there certainly exists 
a coincidence region where 
the decay rates $\Gamma$ in both the numerical 
and analytical results (Eq.(\ref{WKB})) 
coincides (see Fig.7(a)). The best agreement can be seen along the 
coincidence line.
In the vicinity of the coincidence region, the curvatures 
of the left-most and the middle wells are almost the same in 
the original Landau potential $W(x)$, 
and the harmonic levels for each of three wells 
in the effective potential $V(x)$ are 
almost degenerate, satisfying the condition for the 
case (a) below Eq.(\ref{exact-sol}). 
Thus one can see the doubly-humped barrier counterpart 
of Kramers formula to be justified near 
the coincidence line (see Fig.7(b)) 
in the present context.  

In the case of a shallow central well lying 
far from this coincidence region, the degeneracy 
condition is no longer satisfied. At $\lambda=3,\mu=0.3$, 
for instance, the 
middle cave is astonishingly shallow. 
Here, instead of Eq.(\ref{WKB}),
the formula in Eq.(\ref{Kramers}) explains 
the numerical Boltzmann factor.
In Appendix, we alternatively derive the doubly-humped barrier 
counterpart of Kramers rate as the inverse of 
the first passage time. Due to the saddle point approximation, however, 
this method is valid only 
in the case of a shallow central well.  
\begin{figure}[tbp]
\center{
\subfigure{\label{Fig.7(a)}\includegraphics[scale=0.6]{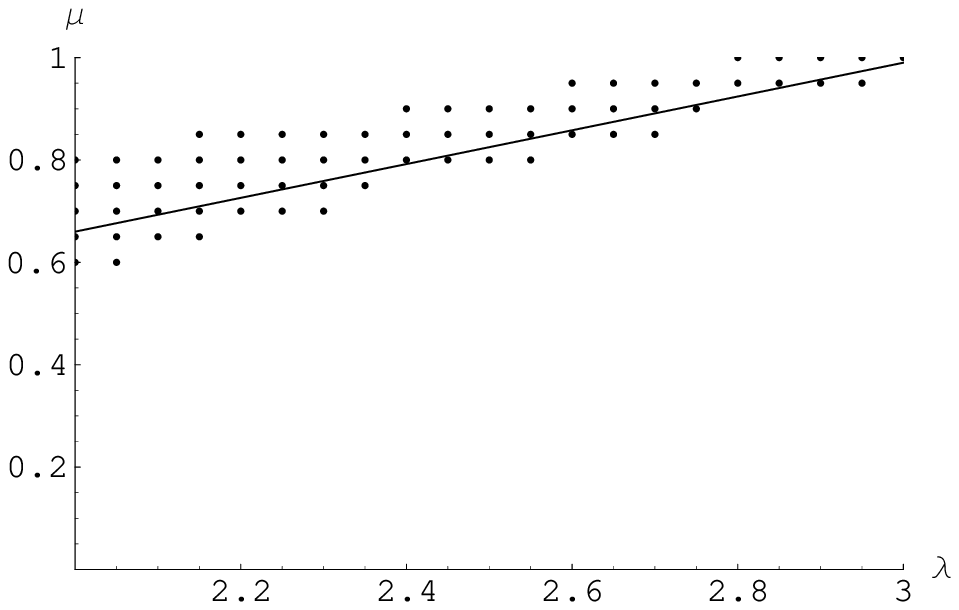}}
\subfigure{\label{Fig.7(b)}\includegraphics[scale=0.6]{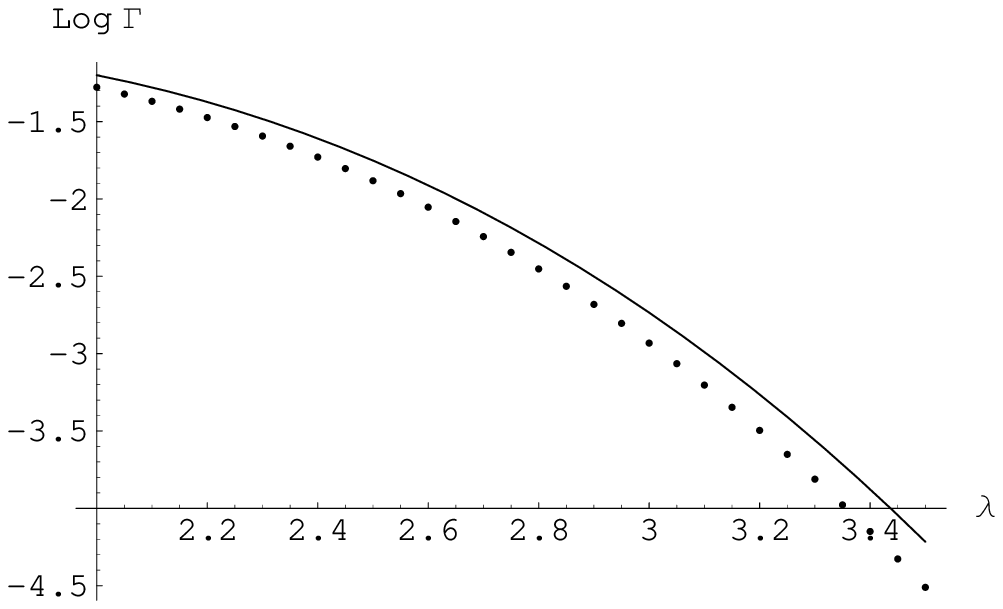}}
}
\caption
{Comparison of the decay rate $\Gamma$ 
between the analytic formula in Eq.(\ref{WKB}) and
numerically obtained values (dark circles) 
in the parameter range $2<\lambda<3$, $0.3<\mu<1$ 
at $\theta=0.1$.
(a)Coincidence region where the ratio 
of the numerical and analytical 
decay rates falls between $0.819$ and $1.221$. 
Along the coincidence line (solid-line), the symmetric level splitting is 
observed as depicted in Fig.(\ref{levelsplit}). 
(b)Comparison between numerical (dark circles)and analytical (solid-line) decay rates 
along the coincidence line in Fig.7(a).}
\end{figure}
\section{\label{summary}Summary and discussions}
With use of
the WKB method, we have analyzed
the homogeneous nucleation phenomenon in systems 
with an intermediate state, 
and obtained
the decay rate for the thermal diffusion over
a doubly-humped barrier.
The analytic result is applicable for a wide range of depth 
and curvature of the intermediate middle well.
In the case of a deep middle well, in particular, 
the decay rate becomes a doubly-humped-barrier version of
the Kramers escape rate:
the barrier height and 
curvature of the initial well in the Kramers rate are now 
replaced by 
the arithmetic mean of higher(or outer) 
and lower(or inner) partial barriers
and the geometric
mean of curvatures of the initial 
and intermediate wells, respectively. 
We have confirmed the presence of  
the coincidence region (almost a narrow strip)
in the parameter space for 
the Landau potential 
where this universal formula holds well, 
and also revealed the condition
for the intermediate state to enhance the nucleation rate.
In the case of 
a shallow middle cave far from
the coincidence parameter region, however, we find 
a less-essential modification of the Kramers rate, 
which is also verified
by an alternative study based on 
the standard framework of 
the mean-first-passage time problem .

\begin{acknowledgments}
TM thanks to professor S.Tasaki for fruitful comments.
This work is supported by JSPS Research Fellowship for young scientists and
21st Century COE Program.
AS and KN are grateful to JSPS for the financial support to the Fundamental
Research.
\end{acknowledgments}
\appendix

\section{Method of 
the mean-first-passage time: the case of a shallow middle cave} 
For a triple-well potential $W$ with doubly-humped barrier, 
the counterpart of Kramers
escape rate is straightforwardly derived with 
the aid of the saddle point approximation for the
mean-first-passage time\cite{Zwanzig}.
The left-most well is approximated as a harmonic potential 
and the middle well 
is given by the fourth-order
symmetric potential, $\frac{a}{2}x^2-\frac{b}{4}x^4$.
The mean first passage time $\tau(x)$  
from an initial point at $x$ is given as
\begin{equation}
\tau(x)=\eta\int_x^{x_a}dy 
e^{\frac{W(y)}{\theta}}\int_{-\infty}^y dz 
e^{-\frac{W(z)}{\theta}},
\end{equation}
where the boundary $x_a$ is larger than two potential maxima. 
We further assume: a Brownian particle can not come back 
once it leaves the domain $[-\infty,x_a]$.
At low temperature, the saddle-point approximation 
for the integral over $z$ is performed as 
\begin{equation}
\int_{-\infty}^y dz e^{-\frac{W(z)}{\theta}}\approx 
\int_{-\infty}^\infty dz 
e^{-\frac{W(z)}{\theta}}
=e^{-\frac{W_{min}}{\theta}}
\sqrt{\frac{2\pi\theta}{\omega_{min}^2}}.
\end{equation}
Here $W_{min}$ and $\omega_{min}$ are the potential minimum 
and the curvature at $x_{min}$ in the left-most well.
\begin{figure}
\center{
\includegraphics[scale=0.5]{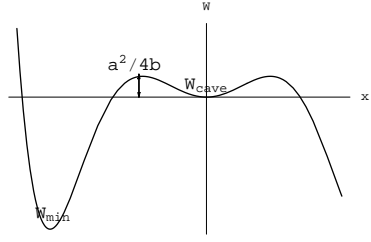}
}
\caption{Schematic doubly-humped potential  with a shallow cave, $W(x)$.}
\label{Figg}
\end{figure}

The integration over $y$ is performed as
\begin{eqnarray}
&&\int_{x}^{x_a} dy e^{\frac{W(y)}{\theta}}
\approx \int_{-\infty}^\infty dy 
e^{\frac{W_{cave}-\frac{a}{2}y^2+
\frac{b}{4}y^4}{\theta}}\nonumber \\
&&=\frac{\pi}{2}\sqrt{\frac{a}{b}}e^{\frac{W_{cave}+
\frac{a^2}{8b}}{\theta}}\left(I_{-\frac{1}{4}}
(\frac{a^2}{8b\theta})+I_{\frac{1}{4}}
(\frac{a^2}{8b\theta})\right) 
\end{eqnarray}
with $W_{cave}$ the potential minimum in the cave .
The decay rate $\Gamma$ is given as the inverse of $\tau$ as
\begin{equation}
\Gamma=\frac{2\omega_{min}}{\pi\sqrt{2\pi 
\theta \frac{a}{b}}}
e^{-\frac{W_{cave}-W_{min}+\frac{a^2}{8b}}
{\theta}}\left(I_{-\frac{1}{4}}
(\frac{a^2}{8b\theta})+I_{\frac{1}{4}}
(\frac{a^2}{8b\theta})\right)^{-1}.
\label{append-4}
\end{equation}
In the low temperature, with use of the 
asymptotic form for the modified Bessel function 
$I_{k}(z)\sim \frac{e^z}{\sqrt{2\pi z}}$, Eq.(\ref{append-4}) 
turns out to be
the usual Kramers escape rate.
\begin{equation}
\Gamma=\frac{\sqrt{a}\omega_{min}}{2\pi}
e^{-\frac{\Delta W}{\theta}},
\end{equation}
where $\Delta W= W_{cave}-W_{min}+\frac{a^2}{4b}$ stands for 
the net barrier height between the minimum of 
the left-most well and the potential top of 
the doubly-humped barrier and $\sqrt{a}$ 
corresponds to the curvature at the top, which 
verifies Eq.(\ref{Kramers}).

\end{document}